\documentclass[prl,reprint]{revtex4-1}
\pdfoutput=1
\usepackage{graphicx}
\usepackage{color}
\usepackage{bm}
\usepackage{hyperref}

\begin{document}

\title{Even-denominator Fractional Quantum Hall Effect at a Landau Level Crossing}
\date{today}

\author{Yang Liu}
\affiliation{Department of Electrical Engineering,
Princeton University, Princeton, New Jersey 08544}
\author{S.\ Hasdemir}
\affiliation{Department of Electrical Engineering,
Princeton University, Princeton, New Jersey 08544}
\author{D.\ Kamburov}
\affiliation{Department of Electrical Engineering,
Princeton University, Princeton, New Jersey 08544}
\author{A.L.\ Graninger}
\affiliation{Department of Electrical Engineering,
Princeton University, Princeton, New Jersey 08544}
\author{M.\ Shayegan}
\affiliation{Department of Electrical Engineering,
Princeton University, Princeton, New Jersey 08544}
\author{L.N.\ Pfeiffer}
\affiliation{Department of Electrical Engineering,
Princeton University, Princeton, New Jersey 08544}
\author{K.W.\ West}
\affiliation{Department of Electrical Engineering,
Princeton University, Princeton, New Jersey 08544}
\author{K.W.\ Baldwin}
\affiliation{Department of Electrical Engineering,
Princeton University, Princeton, New Jersey 08544}
\author{R. Winkler}
\affiliation{Department of Physics,
Northern Illinois University, DeKalb, Illinois 60115}
\affiliation{Materials Science Division, Argonne National Laboratory,
  Argonne, Illinois 60439, USA}

\date{\today}

\begin{abstract}
  The fractional quantum Hall effect (FQHE), observed in
  two-dimensional (2D) charged particles at high magnetic fields, is
  one of the most fascinating, macroscopic manifestations of a
  many-body state stabilized by the strong Coulomb interaction. It
  occurs when the filling factor ($\nu$) of the quantized Landau
  levels (LLs) is a fraction which, with very few exceptions, has an
  odd denominator. In 2D systems with additional degrees of freedom it
  is possible to cause a crossing of the LLs at the Fermi level. At
  and near these crossings, the FQHE states are often weakened or
  destroyed. Here we report the observation of an unusual crossing of
  the two \emph{lowest-energy} LLs in high-mobility GaAs 2D $hole$
  systems which brings to life a new \emph{even-denominator} FQHE at
  $\nu=1/2$.
\end{abstract}

\maketitle

A strong magnetic field $B$ applied perpendicular to the plane of a 2D
electron system quantizes the electron energies into a set of discrete
LLs, separated by the cyclotron energy. With increasing $B$, the
degeneracy of each LL, which is equal to $eB/h$, increases so that the
number of occupied LLs, the filling factor $\nu$, decreases ($e$ is
the electron charge and $h$ is Planck's constant). The discrete LL
structure combined with the decreasing $\nu$ gives rise to the
integral QHE, the formation of incompressible states signaled by a
vanishing longitudinal resistance $R_{xx}$ and a quantized Hall resistance $R_{xy}$, when an integer number of LLs are
fully occupied and the Fermi level ($E_F$) lies in a gap separating
two adjacent LLs \cite{theQHE, PQHE, Jain.CF.2007}. At very low
temperatures, and if disorder is low, electron-electron interaction
leads to new phenomena. An example is the FQHE, the condensation of 2D
electrons into many-body incompressible states which are
phenomenologically similar to the integral QHE, but are stable
predominantly at odd-denominator \emph{fractional} $\nu$ \cite{theQHE,
  PQHE, Jain.CF.2007}.

Now a 2D system can have extra electronic (pseudo-spin) degrees of
freedom, such as spin, valley, layer, or electric subbands. These lead
to additional sets of LLs with different energy separations which can
be tuned, e.g., by tilting the sample in the magnetic field to enhance
the spin Zeeman energy \cite{Poortere.Science.2000}, or by applying
uniaxial strain to manipulate the valley splitting energy
\cite{Gunawan.PRL.2006}. The tuning causes the LLs to cross each
other. When a crossing occurs at $E_F$, in a single-electron picture
there is no gap at $E_F$ and the QHE is destroyed. In an interacting
system, however, depending on the pseudo-spins of the crossing LLs,
one can have QHE \emph{ferromagnetism} where a gap opens at the
crossing and stabilizes a ferromagnetic QHE state
\cite{Jungwirth.PRB.2000, Muraki.PRL.2001}. For example in a
two-valley system, even when the LLs of the valleys are tuned to have
the same energy, both the integral QHE at $\nu=1$ and the FQHE
$\nu=1/3$ survive \cite{Shkolnikov.PRL.2005, Padmanabhan.PRL.2010}. In
other systems, when LLs belonging to different electric subbands and
with opposite spins cross, the FQHE disappear \cite{Liu.PRB.2011,
  Liu.PRL.2011B}.

Here we describe unexpected phenomena in 2D hole systems (2DHSs)
confined to GaAs quantum wells (QWs). We observe an unusual crossing
of the two lowest-energy LLs. The crossing leads to a
weakening or disappearance of the commonly seen odd-denominator FQH
states in the filling range $1/3\le\nu\le 2/3$. But, surprisingly, a
new FQH state at the \emph{even-denominator} filling $\nu=1/2$ comes
to exist at the crossing.

Our samples were grown on GaAs (001) wafers by molecular beam
epitaxy. The 2DHS is confined to 30- and 35-nm-wide GaAs QWs, flanked
by undoped Al$_{0.3}$Ga$_{0.7}$As spacer layers and carbon
$\delta$-doped layers, and have a very high mobility $\mu \geq$ 200
m$^2$/Vs at low temperatures.  We made samples in a van der Pauw
geometry, 4 $\times$ 4 mm$^2$, and alloyed In:Zn contacts at their
four corners. We then fitted each sample with an evaporated Ti/Au
front-gate and an In back-gate to control the symmetry of the charge
distribution in the QW and 2D hole density, $p$, which we give
throughout this paper in units of $10^{11}$ cm$^{-2}$. Unless
otherwise noted, all the data presented here were taken in QWs with
symmetric charge distributions. The measurements were carried out in a
dilution refrigerator with a base temperature of $T \approx$ 30 mK and
a superconducting magnet up to 18 T.

\begin{figure*}
\includegraphics[width=.8\textwidth]{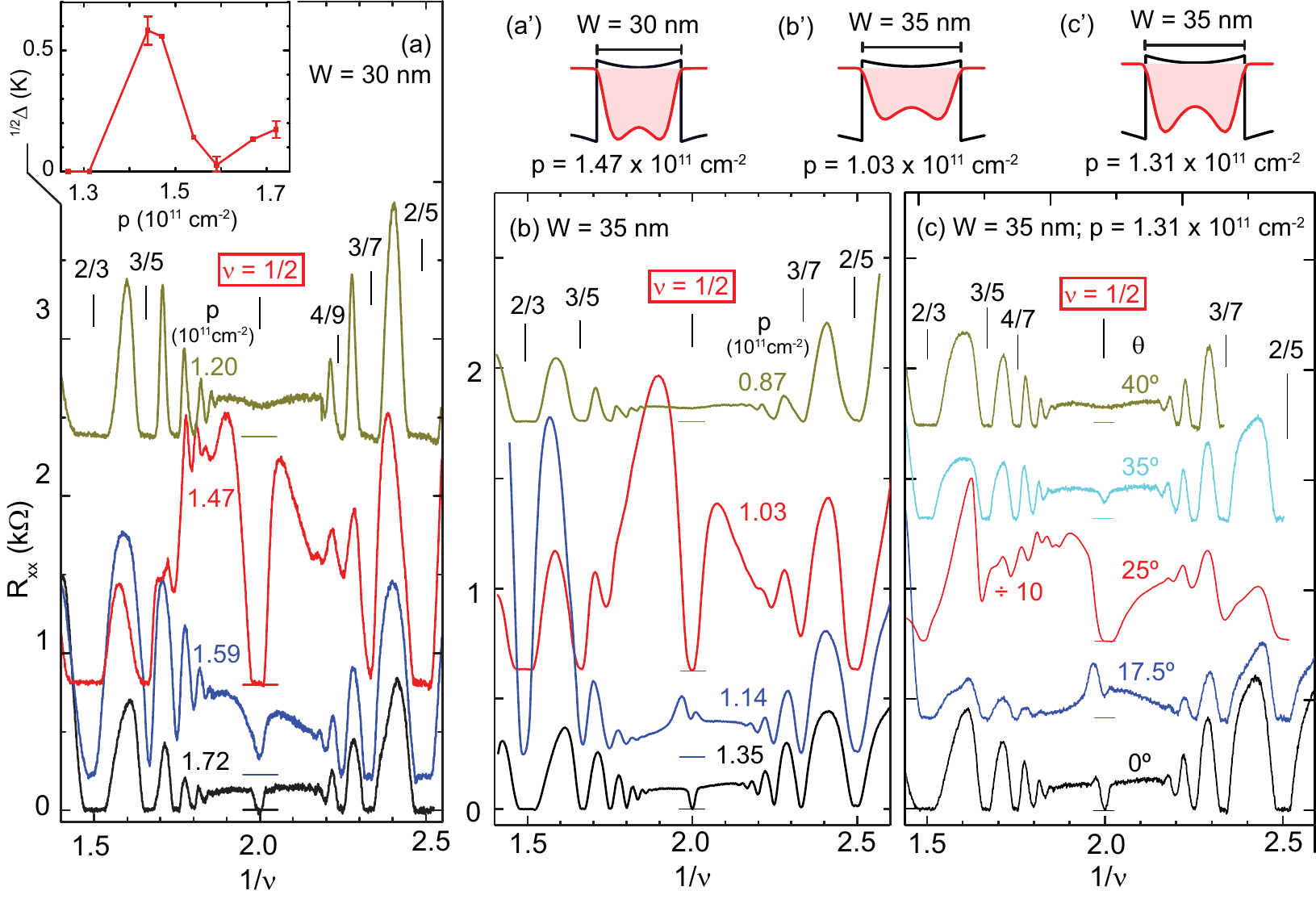}
\caption{(a) \& (b): Magneto-resistance ($R_{xx}$) traces for 2D holes
  confined to 30- and 35-nm-wide GaAs QWs at several densities. The inset in (a) is the measured $\nu=1/2$ FQH energy
  gap. (c)
  $R_{xx}$ for 2D holes confined to the 35-nm-wide QW at density
  $p=1.31\times 10^{11}$ cm${^{-2}}$ and different tilting angles, as indicated. The traces in
  (a)-(c) are shifted vertically for clarity. (a') - (c'): Results of self-consistently calculated (at $B=0$) charge distribution (red curve) and potential (black curve).}
\end{figure*}


\begin{figure}
\includegraphics[width=.48\textwidth]{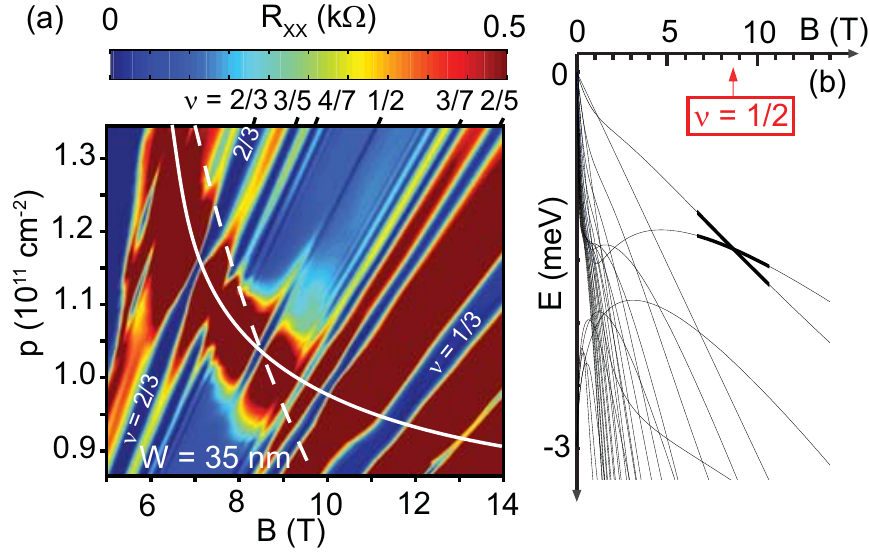}%
\caption{(a) A Color-scale plot of ${R_{xx}}$ vs $B$ for the
  35-nm-wide QW as $p$ is changed from $0.85$ to $1.35\times 10^{11}$
  cm${^{-2}}$. The solid white curve is a guide to the eye, showing
  the observed LL crossing trajectory. The dashed white curve
  represents the calculated position of the high-$B$ LL crossing (see
  text). (b) Calculated LL fan diagram, showing a crossing
  of the two lowest-energy levels at high $B$ ($\approx 8.6$ T).}
\end{figure}


\begin{figure}[htb]
\includegraphics[width=.48\textwidth]{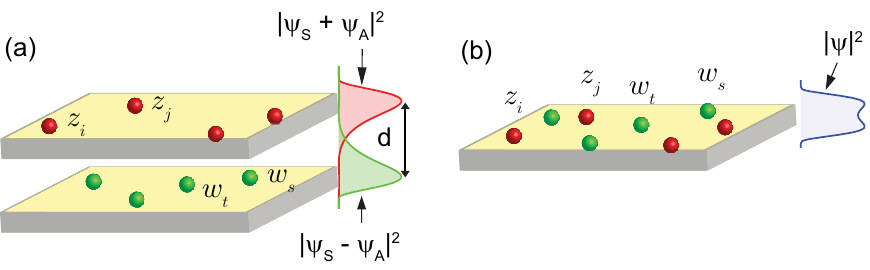}%
\caption{Schematic figures showing: (a) $\Psi_{331}$
  state with layer degree of freedom. The two layers are linear
  combinations of the symmetric ($\psi_S$) and antisymmetric
  ($\psi_A$) subbands, and the total charge distribution is bilayer-like. The
  state has $\nu=1/3$ FQHE-like correlations in each layer, as well as
  inter-layer correlation (Eq. 1). (b) For the hole systems in our experiments,
  the charge distribution is essentially single-layer-like so that the two
  pseudo-spin species effectively reside in the same layer.}
\end{figure}

Figures 1-3 capture the highlights of our work. In Fig. 1(a) we show
$R_{xx}$ traces for 2D holes confined to the 30-nm-wide-QW for several
densities, from 1.20 to 1.72 \cite{Note1}. 
The $x$-axis in Fig. 1(a)
is the $\nu^{-1}$ in order to normalize $B$ and align the FQH states
for the different traces. At the lowest $p$ (top trace), $R_{xx}$
minima are observed at numerous fillings such as $\nu=$ 2/3, 3/5, 4/7,
..., and 2/5, 3/7, 4/9, ..., attesting to the very high quality of the
sample. These FQH states and their relative strengths resemble those
typically seen in high-mobility 2D holes (or electrons) confined to
GaAs QWs. More important, they are all at odd-denominator $\nu$ and
there is no FQHE at $\nu=1/2$. As $p$ increases, starting at $p\approx
1.32$ \cite{Note1}, an $R_{xx}$ minimum develops at $\nu$ = 1/2, and
quickly deepens and turns into a zero-resistance plateau centered at
$\nu=1/2$ for $p=1.47$. Concomitantly, the $R_{xy}$ trace
\cite{Note1} exhibits a Hall plateau quantized at $2h/e^{2}$,
signaling the formation of a strong FQH state at $\nu=1/2$. At a
slightly higher density, $p=1.59$, the $\nu=1/2$ $R_{xx}$ minimum
becomes weak, but returns again at higher $p$ and the $\nu=1/2$ FQHE
persists up to the highest densities we can achieve in this
sample. Data for another QW sample with a wider well width of 35 nm,
shown in Fig. 1(b), exhibit a qualitatively similar evolution, albeit
in a somewhat lower density range. In particular, the $\nu=1/2$ FQHE
is strongest at $p=$ 1.03.

The behavior of $R_{xx}$ near $\nu=1/2$, and in particular the
strengths of the nearby FQH states, provide clear hints that a LL
crossing is occurring in Figs. 1(a,b) data sets. This crossing is
better seen in the color-coded plot of Fig. 2(a), which condenses many
$R_{xx}$ traces for the 35-nm-QW sample (see \cite{Note1} for a
similar plot of the 30-nm-QW data). On the flanks of $\nu=1/2$ and for
$0.9 < p < 1.1$, there is an overall rise of $R_{xx}$, seen as a red
band in Fig. 2(a) which starts at low $p$ on the right side (high
field side) of $\nu=1/2$ and gradually moves to the left and past
$\nu=1/2$ with increasing $p$. Away from this band, the
odd-denominator FQH states are strong, indicated by deep $R_{xx}$
minima seen as blue stripes. The solid white line in Fig. 2(a) is a
guide to the eye and shows the trajectory of the LL crossing, when the
odd-denominator FQH states become weak or disappear. Surprisingly, the
$\nu=1/2$ FQHE is very strong at this crossing.

The data presented so far indicate a crossing near $\nu=1/2$ between
the two lowest-energy LLs of the 2DHS. But how could there be such a
crossing? In most 2D systems, e.g., \textit{electrons} confined to a
GaAs QW, the energy separation between the lowest two LLs at a given
$\nu$ is determined by either the Zeeman or cyclotron energies, both
of which increase with density, or by the electric subband separation
(in sufficiently wide QWs) which decreases with increasing density
\cite{Suen.PRL.1994} but this separation remains finite and the two
lowest-energy LLs do not cross. However, in 2D $hole$ systems confined
to relatively wide GaAs QWs where the heavy-hole (HH) and light-hole
(LH) subbands are close in energy, an unusual crossing of the two
lowest-energy LLs does indeed occur. This crossing can be
qualitatively understood in a simple picture. The HH and LH bands
have, respectively, heavy and light effective masses for the (bound)
out-of-plane motion. For the in-plane motion, however, the HH bands
have a smaller mass compared to the LH bands
\cite{Winkler.SOC.2003}. The energy of the lowest LL of the HH bands
therefore increases more rapidly with $B$, and eventually intersects
the lowest LL of the LH bands at sufficiently high $B$.

In a more quantitative picture, besides crossings, there is
significant mixing and repulsion between the LLs in a 2DHS so that a
more complex, non-linear LL fan diagram ensues. We demonstrate this in
Fig. 2(b) where we show the results of our calculations, based on an
$8\times 8$ Kane model \cite{Winkler.SOC.2003}, of the LL fan diagram
for a 2DHS at $p=1.05$ and confined to a 25-nm-wide GaAs QW \cite{Note2}.
At $B=0$, the two lowest-energy subbands are HH-like. The
application of $B$ leads to mixing and numerous crossings and
anti-crossings between various LLs of the HH and LH subbands, as seen
in Fig. 2(b). Of particular interest to us is the crossing between the
two lowest-energy levels at $B\approx8.6$ T which corresponds to
$\nu=1/2$ for $p=1.05$ \cite{Note3, Note4}. 
We believe this is the
crossing at which the $\nu=1/2$ FQHE emerges in our samples.

Three qualitative features of this LL crossing favor our
interpretation. First, for a fixed QW width, the calculated field
position of the LL crossing moves from high $B$ to low $B$ as $p$ is
increased; see the dashed white curve in Fig. 2(a). The calculated
curve has a steeper dependence on $p$ compared to the crossing seen
experimentally (solid white curve), but the behavior is qualitatively
similar and explains the successive weakening or disappearance and
reappearance of the nearby, odd-denominator FQH states as $p$ is
varied. Second, our calculations show that for a narrower QW, the
crossing at $\nu=1/2$ moves to a larger $p$, again consistent with the
experimental observations in Figs. 1(a,b); also, compare Fig. 2(a)
with Fig. S2 of \cite{Note1}. Third, for a fixed QW width and $p$,
our calculations indicate that when the charge distribution in the QW
is made asymmetric, the crossing turns into an anti-crossing, meaning
that the two lowest-energy LLs are always separated by a finite energy
gap. This is also consistent with our experimental data: Even for a
very small ($\leq 5\%$) charge distribution asymmetry in the QW, which
we induce by adjusting the front- and back-gate biases, the nearby FQH
states no longer exhibit a pronounced weakening, consistent with the
absence of a LL crossing, and the $\nu=1/2$ FQHE disappears.

Having established that the $\nu=1/2$ FQH state we observe indeed
emerges at a LL crossing, it is natural to interpret it as a
two-component FQH state described by the Halperin-Laughlin
$\Psi_{331}$ wavefunction \cite{Halperin.HPA.1983}:
\begin{equation}
  \Psi_{331} = \prod_{i,j} (z_i-z_j)^3 \prod_{s,t}(w_s-w_t)^3  \prod_{i,s}(z_i-w_s)^1,
\end{equation}
\noindent which has strong correlations between the two pseudo-spin
components ($z$ and $w$ denote the coordinates of electrons with
different pseudo-spins). The $\Psi_{331}$ state is believed to
describe the $\nu=1/2$ FQHE seen in $bilayer$ electron systems
\cite{Suen.PRL.1992, Eisenstein.PRL.1992, Suen.PRL.1994,
  Shabani.PRB.2013, Yoshioka.PRB.1989, He.PRB.1993,
  Papic.PRB.2009a, Papic.PRB.2009, Peterson.PRB.2010,
  Peterson.PRB.2010b, Note5} 
confined to either wide GaAs QWs
\cite{Suen.PRL.1992, Suen.PRL.1994, Shabani.PRB.2013} or to
double-QWs \cite{Eisenstein.PRL.1992, Note5}.  In these systems, the
two components are the "layers" which can be constructed through
linear combinations of the symmetric and antisymmetric states. In
electron systems confined to wide GaAs QWs, the energy separation
between the states can be substantial but if it becomes only a small
fraction ($\lesssim 0.1$) of the in-plane Coulomb energy $E_C=e{^2}/ 4
\pi \epsilon l_{B}$, a $\nu=1/2$ FQHE emerges ($\epsilon$ is the GaAs
dielectric constant and $l_B$ the magnetic length)
\cite{Suen.PRL.1994, Shabani.PRB.2013}. Returning to the 2DHSs in
our study, it is natural to associate the two crossing LLs with the
two pseudo-spins needed for a $\Psi_{331}$ state. Since the energy
separation between these two LLs is zero at the crossing, $E_C$
certainly dominates over the pseudo-spin energy separation. The
emergence of a strong $\nu=1/2$ FQHE at the crossing is therefore
plausible.

The above interpretation raises interesting questions. First, in the
case of electrons which are confined to either a wide or double GaAs
QWs, the $\nu=1/2$ FQHE is observed only when the charge distribution
is bilayer-like (Fig. 3(a)) \cite{Suen.PRL.1992, Eisenstein.PRL.1992,
  Suen.PRL.1994, Shabani.PRB.2013}. In our hole system, however,
the charge distribution is essentially single-layer-like at $B=0$
(Fig. 3(b); see also Figs. 1(a',b')). Although we do not know the
exact charge distribution at $\nu=1/2$, we expect it to be
qualitatively similar to $B=0$ because deviations would cost
significant electrostatic energy. Moreover, the mixing of LH states
\cite{Note4}, which have smaller out-of-plane mass, should favor a
single-layer-like charge distribution. It appears then that we are
observing a $\Psi_{331}$ FQH state in an essentially single-layer
system (Fig. 3(b)). This is very surprising. Although the $\Psi_{331}$
state was originally proposed for a 2D system with spin degree of
freedom \cite{Halperin.HPA.1983}, a $\nu=1/2$ FQHE has never been
reported for systems with spin or valley
\cite{Shayegan.Phys.Stat.Sol.b.2006} degrees of freedom where the
particles with different pseudo-spins are essentially in the same
layer \cite{Note6, Note7}.

Second, is there a $\nu=1/2$ FQHE in 2DHSs which is equivalent to the
state seen in electron systems in wide GaAs QWs, i.e., is stabilized
when the system is bilayer-like? We believe that the $\nu=1/2$ FQHE we
observe at high $p$, well past the LL crossing, is indeed such a
state. As reported elsewhere \cite{Liu.PRL.2014}, at very high
$p$, the 2DHS has a bilayer charge distribution which stabilizes a
$\nu=1/2$ FQHE. But the data in Figs. 1 and 2 for both samples show
that, as $p$ is increased so that the LL crossing is moved to $\nu$
larger than $1/2$, the $\nu=1/2$ FQHE \emph{weakens} and then
\emph{reappears} again at higher $p$. The weakening is readily seen as
a rise in $R_{xx}$ from zero at $p=1.59$ for the 30-nm and at $p=1.14$
in the 35-nm-QW samples [see Figs. 1(a,b) and 2(a)]. For the
30-nm-wide QW, we also measured the $\nu=1/2$ FQHE energy gap
$^{1/2}\Delta$ from the Arrhenius plots of $R_{xx}$ minimum vs inverse
temperature, for several densities. The data, shown in Fig. 1(a)
inset, corroborate our conclusion: $^{1/2}\Delta$ is highest at the LL
crossing ($p\approx 1.45$) and exhibits a minimum at $p\approx 1.60$
before becoming larger again at higher $p$. This weakening signals a
transition between a $\nu=1/2$ FQHE stabilized by the LL crossing, and
one stabilized by the bilayer-like charge distribution at high $p$.

Finally, we present results from our preliminary study of the same
2DHSs in the presence of an additional parallel magnetic field
($B_{||}$) applied in the 2D plane; we introduce this $B_{||}$ by
tilting the sample so that the total applied field makes an angle
$\theta$ with respect to the normal to the 2D plane. The data for the
35-nm-QW at a fixed density of $p=1.31$ are shown in Fig. 1(c). Note
that this density is higher than $p=1.03$ at which the two
lowest-energy LLs cross at $\nu=1/2$ at $\theta=0$ (see Figs. 1(b) and
2(a)). Consistent with Figs. 1(b) and 2(a), we indeed observe a
reasonably well-developed $\nu=1/2$ FQHE at $\theta=0$ in
Fig. 1(c). As we tilt the sample, the $\nu=1/2$ FQHE becomes weaker
and in fact disappears at $\theta=17.5^{\circ}$. Surprisingly, at
$\theta=25^{\circ}$, a very strong FQHE reappears at $\nu=1/2$. Note
that in the $\theta=17.5^{\circ}$ and $25^{\circ}$ traces in
Fig. 1(c), the surrounding odd-denominator FQHEs become weak,
signaling a LL crossing near $\nu=1/2$. As the sample is further
tilted to higher $\theta$, the $\nu=1/2$ FQHE becomes weak and
disappears, while the other FQH states get strong. The trace at
highest angle, $\theta=40^{\circ}$, indeed has a strong resemblance to
the trace taken at $low$ $p$ in this sample at $\theta=0$ (see
Fig. 1(b) top trace).

The evolution in Fig. 1(c) as a function of $\theta$ is very different
from what is seen for 2D $electrons$ confined to wide GaAs QWs. There,
the $\nu=1/2$ FQHE either quickly disappears as the sample is tilted,
or initially becomes slightly stronger before disappearing
\cite{Suen.PRL.1992, Lay.PRB.1997, Luhman.PRL.2008,
  Liu.PRL.2014}. On the other hand, the evolution in Fig. 1(c) is
qualitatively similar to what we observe in the 2DHSs at $\theta=0$ as
we decrease $p$ (see Figs. 1(a,b)), suggesting that $B_{||}$ induces a
LL crossing near $\nu=1/2$. This is not unexpected. Previous
measurements in similar 2DHSs have indeed revealed a crossing of the
two lowest-energy LLs as the sample is tilted in $B$
\cite{Lewis.PRL.2011}. In Fig. 1(c) it is particularly noteworthy that
the $\nu=1/2$ FQHE essentially disappears (at $\theta=17.5^{\circ}$)
before becoming very strong at the LL crossing
($\theta=25^{\circ}$). This is similar to what happens at $\theta=0$
at $p=1.14$ in the same sample (Fig. 1(b)), providing further evidence
that a compressible state appears to separate the $\nu=1/2$ FQHE
observed at high $p$ from the one seen at the LL crossing.

It is remarkable that the LL crossing induced by
$B_{||}$ causes a significant increase in $R_{xx}$ near
$\nu=1/2$. Note that the $\theta=25^{\circ}$ trace in Fig. 1(c) is shown
reduced by a factor of 10, meaning that although $R_{xx}$ at
$\nu=2/3$, 1/2, and 2/5 is close to zero, $R_{xx}$ at other fields far
exceeds the values in traces taken at other angles. This rise in
$R_{xx}$ is more pronounced compared to the rise seen near the LL crossing
at $\theta=0$ (Figs. 1(a,b)). Evidently, the addition of $B_{||}$
adds yet another twist to the fascinating phenomena observed in
clean, interacting 2DHSs.

The results presented here attest to the rich many-body physics of the
2DHSs confined to GaAs QWs. They demonstrate an unusual crossing of
the two lowest-energy hole energy levels in a large perpendicular magnetic field and,
more remarkably, the emergence of a very rare,
\textit{even-denominator} FQHE at the crossing.
We have tentatively interpreted this FQHE as a two-component
$\Psi_{331}$ state, although a detailed and quantitative understanding
of its origin and properties await future research.

\begin{acknowledgments}
  We acknowledge support through the DOE BES (DE-FG02-00-ER45841) for
  measurements, and the Gordon and Betty Moore Foundation (Grant
  GBMF2719), Keck Foundation, and the NSF (DMR-1305691 and MRSEC
  DMR-0819860) for sample fabrication. Work at Argonne was supported
  by DOE BES (DE-AC02-06CH11357). Our work was partly performed at the
  National High Magnetic Field Laboratory, which is supported by NSF
  (DMR-1157490), the State of Florida, and the DOE. We thank
  J. K. Jain and Z. Papic for discussions, and S. Hannahs, E. Palm,
  J. H. Park, T. P. Murphy, G. E. Jones, and A. Suslov for technical
  assistance.
\end{acknowledgments}

\section{Supplementary Materials}

In this Supplement, we present additional magneto-transport data for
the two-dimensional (2D) holes confined to the 30-nm-wide GaAs quantum
well sample as the 2D hole density ($p$) is changed, while the charge
distribution is kept symmetric. In the lower part of Fig. S1 we show
$R_{xx}$ traces for several densities, ranging from 1.20 to 1.72 (all
densities are given in units of $10^{11}$ cm$^{-2}$). At the lowest
$p$ (top $R_{xx}$ trace), minima in $R_{xx}$ are observed at numerous
odd-denominator fillings such as $\nu=$ 2/3, 3/5, 4/7, ..., and 2/5,
3/7, 4/9, ..., signaling strong fractional quantum Hall effect (FQHE)
states. As $p$ increases, $R_{xx}$ develops a minimum at $\nu$ = 1/2,
starting at $p\approx 1.32$. This minimum quickly deepens and turns
into a zero-resistance plateau centered at $\nu=1/2$ for
$p=1.47$. Concomitantly, the $R_{xy}$ trace, shown in the upper part
of Fig. S1, exhibits a Hall plateau quantized at $2h/e^{2}$, providing
evidence for the formation of a strong FQH state at $\nu=1/2$. At
slightly higher densities ($p=1.54$ and 1.59), the $\nu=1/2$ $R_{xx}$
minimum becomes weaker, but then it turns deep again and the FQHE persists up to the highest
densities we can achieve in this sample.

\begin{figure}
\includegraphics[width=.41\textwidth]{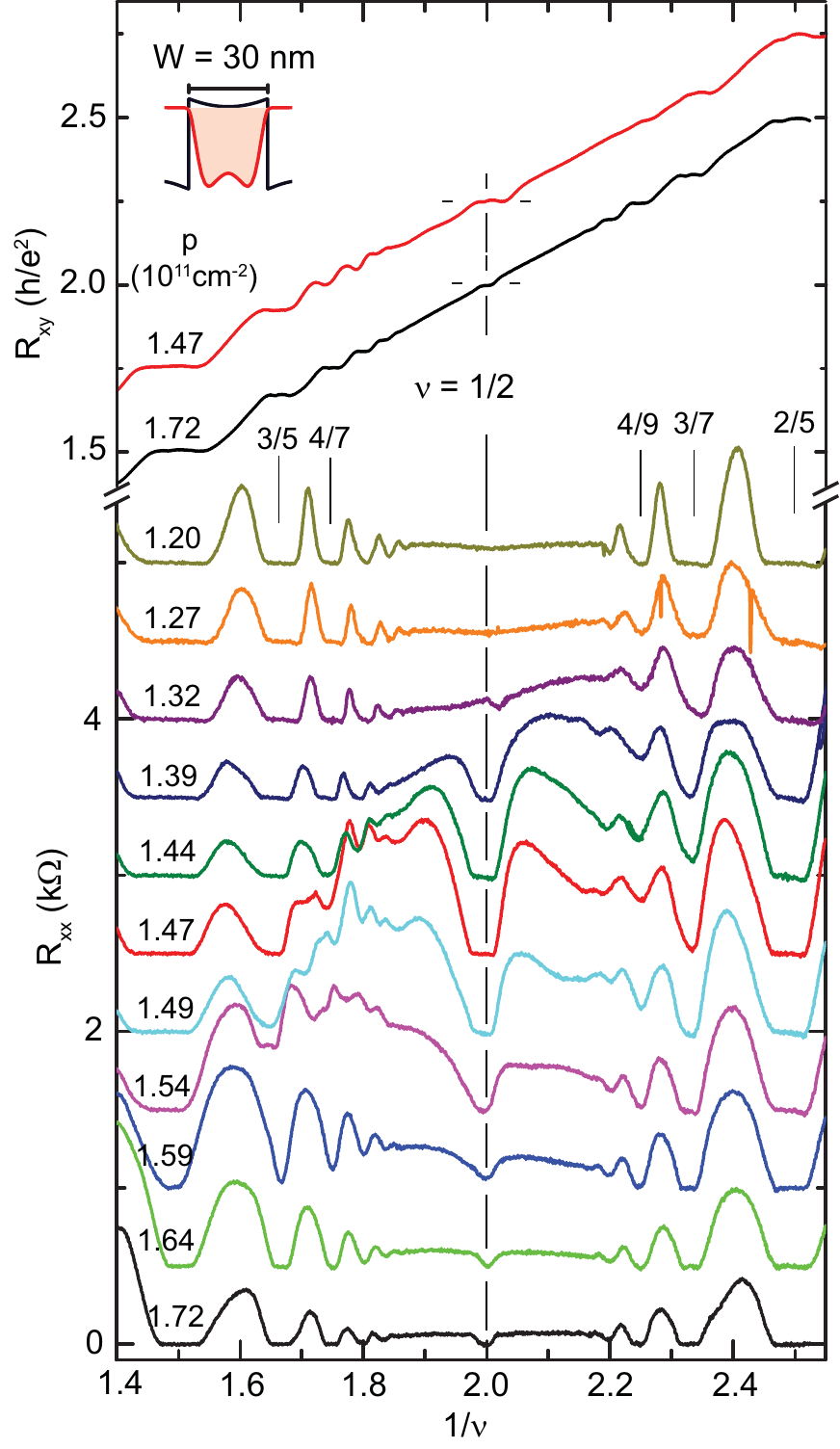}
\caption{(Fig. S1)$R_{xx}$ and $R_{xy}$ magneto-resistance traces for 2D holes
  confined to a 30-nm-wide GaAs quantum well. The density for each trace is
  indicated (in units of $10^{11}$ cm${^{-2}}$), and traces are
  shifted vertically for clarity. The top inset shows the ($B=0$)
  calculated charge distribution (red curve) and potential (black
  curve) in this sample at $p=1.47\times 10^{11}$ cm${^{-2}}$.}
\end{figure}

In Fig. S1, note the rise in $R_{xx}$ on the right (high field) side
of $\nu=1/2$ as the density is raised from $p=1.27$ and 1.32. The rise is seen
nearly symmetrically on both sides of $\nu=1/2$ at $p=1.44$ and 1.47, and then
it moves to the left of $\nu=1/2$ as density is further increased; e.g., see the $p=1.54$ trace. At the highest densities (lowest two
trances) the $R_{xx}$ rise has disappeared and the traces look
qualitatively similar to the low-density trace except that now there
is a strong FQHE at $\nu=1/2$. The rise in $R_{xx}$ qualitatively
resembles a "wave" that moves from right to left (of $\nu=1/2$) as the
density is raised and, in its passage, weakens or destroys the
odd-denominator FQH states. We associate the rise in $R_{xx}$ with a
crossing of the two lowest-energy Landau levels.

\begin{figure}
\includegraphics[width=.48\textwidth]{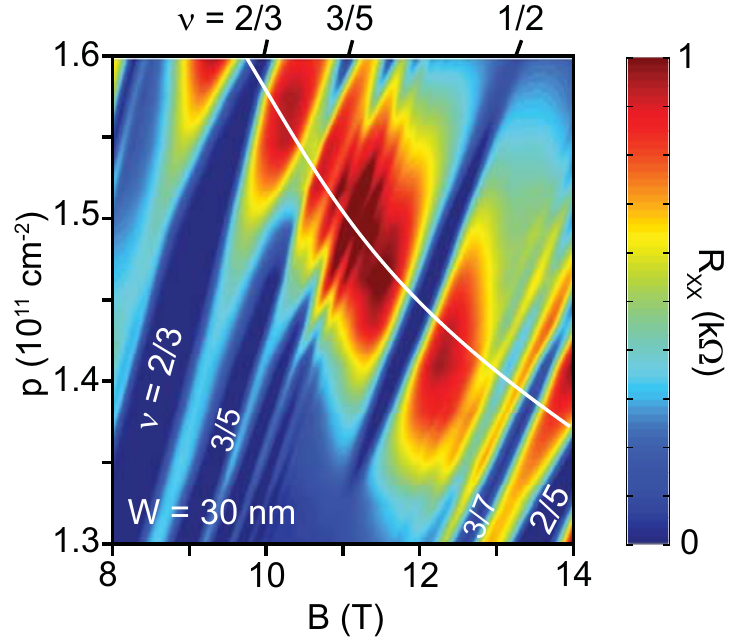}
\caption{(Fig. S2) A color-scale plot of ${R_{xx}}$ vs magnetic field for
  the 30-nm-wide GaAs quantum well as $p$ is changed from $1.3$ to $1.6\times
  10^{11}$ cm${^{-2}}$. The red band shows the trajectory of the Landau Level
  crossing as a function of $p$, and the white curve is a guide to the
  eye.}
\end{figure}

For a better view of this crossing we provide the color-coded plot of
Fig. S2. The overall rise of $R_{xx}$ is seen as a red band, which
starts at low $p$ on the right side of $\nu=1/2$ and gradually moves
to the left and past $\nu=1/2$ with increasing $p$. Away from this
band, the odd-denominator FQH states are strong, indicated by deep
$R_{xx}$ minima seen as blue stripes. The white line in Fig. S2 is a
guide to the eye and shows the trajectory of the Landau Level
crossing, which causes the odd-denominator FQH states to become weak or
disappear. Clearly, the $\nu=1/2$ FQH state is very strong at this
crossing, and becomes weaker on either side of the crossing.

\bibliographystyle{apsrev4-1}
\bibliography{../bib_full}

\end{document}